\documentclass[conference]{IEEEtran}
\IEEEoverridecommandlockouts
\usepackage{cite}
\usepackage{amsmath,amssymb,amsfonts}
\usepackage{algorithmic}
\usepackage{graphicx}
\usepackage{textcomp}
\usepackage{xcolor}
\def\BibTeX{{\rm B\kern-.05em{\sc i\kern-.025em b}\kern-.08em
    T\kern-.1667em\lower.7ex\hbox{E}\kern-.125emX}}
    
\newcommand{\ignore}[1]{}
\newcommand{\A}{\mathbb A}
\newcommand{\sA}{\mathbb A}
\newcommand{\B}{\mathbb B}
\newcommand{\sB}{\mathbb B}

\newcommand{\Q}{\mathbb Q}

\newcommand{\G}{\mathbb G}
\newcommand{\sH}{\mathbb H}
\newcommand{\sP}{\mathbb P}

\DeclareMathOperator{\CSP}{CSP}
\DeclareMathOperator{\Pol}{Pol}

\newtheorem{theorem}{Theorem}[section]

\newtheorem{definition}[theorem]{Definition}

\newtheorem{example}[theorem]{Example}
\newtheorem{conjecture}[theorem]{Conjecture}

\begin{document}

\title{Current Challenges in\\ Infinite-Domain Constraint Satisfaction:\\  Dilemmas of the Infinite Sheep
\thanks{Michael Pinsker has received funding from the Austrian Science Fund (FWF) through project No P32337. The two yet unpublished examples we present in Section~\ref{sect:blackboxes} are from joint work with A.~Mottet and T.~Nagy. The dilemmas are not presented in the  order expected.}
}

\author{\IEEEauthorblockN{Michael Pinsker}
\IEEEauthorblockA{\textit{Institute of Discrete Mathematics and Geometry} \\
\textit{Technische Universit\"{a}t Wien}\\
Vienna, Austria \\
Email: michael.pinsker@tuwien.ac.at, ORCID:  0000-0002-4727-918X}
}

\maketitle

\begin{abstract}
A Constraint Satisfaction Problem (CSP) is a computational problem
where we are given variables and constraints about them; the
question is whether the variables can be assigned values such that all constraints are satisfied. 
We give an overview
of the current state of research on CSPs where values for the variables and constraints are taken  from a finitely bounded homogeneous structure which is fixed beforehand. We explain  the main
mathematical ideas so far, the three dilemmas they brought upon us, and what could  be done to overcome them in
order to obtain a satisfactory understanding of the computational complexity of such CSPs.
\end{abstract}

\begin{IEEEkeywords}
constraint satisfaction, $\omega$-categoricity, finitely bounded homogeneous structure,  canonical function, Ramsey structure, local consistency,  sheep
\end{IEEEkeywords}

\section{Introduction}

In a \emph{Constraint Satisfaction Problem (CSP)}, one is given a finite set $V$ of \emph{variables}  and a finite set $C$ of \emph{constraints} about them and has to decide whether the variables in $V$ can be assigned \emph{values} such that all constraints in $C$ are satisfied. In this article, we only consider \emph{fixed-template} CSPs, in which case there is a set $A$ (called the \emph{domain}) of possible values the variables may take which is fixed beforehand, and where the constraints in $C$ are taken from a fixed finite set of relations on $A$. The problem is thus determined by a relational structure $\A=(A; R_1^\A,\ldots,R_m^\A)$, called a  \emph{template} of the CSP, which we then  denote by $\CSP(\A)$. As an \emph{instance} $I$ of the problem, we are then given a set of variables, say $V:=\{x_1,\ldots,x_6\}$, and a list of constraints given as atomic formulas over $\A$, say $R_1(x_1,x_1,x_2), R_1(x_2,x_3,x_5), R_2(x_2,x_5)$. The question is whether there is a \emph{solution} to this instance $I$, which is a map $s\colon V\to A$ making all constraints true; in our example, a map  such that  $R_1^\A(s(x_1),s(x_1),s(x_2))$, $R_1^\A(s(x_2),s(x_3),s(x_5))$, $R_2^\A(s(x_2),s(x_5))$ all hold in $\sA$. Research on CSPs generally aims at relating structural properties of the template $\A$ to  the computational complexity of the CSP it defines.

\begin{example}\label{ex:diophantine}
If $\A$ is the template whose domain is the set $\mathbb Z$ of integers,  and which has the unary relations $\{0\}$ and $\{1\}$, as well as the ternary relations $\{(x,y,z)\;|\; x+y=z\}$ and $\{(x,y,z)\;|\; x\cdot y=z\}$, then $\CSP(\mathbb A)$ essentially amounts to the problem of deciding whether a given finite set of arithmetic equations has a solution in $\mathbb Z$. This is undecidable~\cite{Matijasevic9}. Note that here we simply encode the ring of the integers as a relational structure to fit our definition of a CSP; for research on  templates with functions see~\cite{functionalCSPs}.
\end{example}

\begin{example}\label{ex:linear}
If $\A$ is the template whose domain is the domain of a finite field $\mathbb F=(F;0,1,+,\cdot)$, and which has the unary relations $\{0\}$ and $\{1\}$ and the  ternary relation $\{(x,y,z)\;|\; x+y=z\}$, then $\CSP(\sA)$ is essentially the problem of deciding whether a given finite set of linear equations has a solution in $\mathbb F$. This problem is in P, i.e., solvable in polynomial time in the number of variables.
\end{example}

\begin{example}\label{ex:3-SAT}
If $\A$ has the Boolean domain $\{0,1\}$, and a single ternary relation $\{(1,0,0),(0,1,0),(0,0,1)\}$, then $\CSP(\sA)$ is the 1-in-3-SAT problem of deciding whether we can assign Boolean values to the variables so that specified triples contain precisely one $1$ -- an NP-complete problem.   
If instead we equip $\sA$ with all binary relations on the Boolean domain, we obtain the problem 2-SAT, which is in P.
\end{example}

\begin{example}\label{ex:acyclic}
If $\A$ is the template whose domain is the set $\mathbb Q$ of rational numbers, and whose only relation is the usual strict order $<^\Q$ on them, then $\CSP(\sA)$ is essentially the problem of deciding whether a given directed graph is acyclic. This problem is in P.
\end{example}

\begin{example}\label{ex:3coloring}
If $\A$ has domain $\{0,1,2\}$, and has as its only relation the disequality relation $\neq^{\{0,1,2\}}$ on this set, then $\CSP(\sA)$ is essentially the problem of deciding whether a given graph is 3-colorable, which is NP-complete.
\end{example}

Note that while every instance $I$ of a CSP, i.e., the variable set $V$ and the list $C$ of constraints given by atomic formulas, is necessarily finite for the problem to be of  computational nature, it is in general  neither necessary nor natural that the domain of the template, i.e., the set of possible values the variables can take, be finite as well. However, for finite-domain templates, a strong algebraic theory 
was developed which not only  culminated in the celebrated confirmation of the Feder-Vardi conjecture from~\cite{FederVardi} by Bulatov~\cite{BulatovFVConjecture} and Zhuk~\cite{Zhuk20}  stating that in this case, the CSP is always either in P or NP-complete, but also provided a clear structural dividing line between those cases (assuming P$\neq$NP, which we shall henceforth do). Before that, the theory moreover led to the characterization of those templates whose CSP has \emph{bounded width}, i.e.,  can be correctly solved by  checking local consistency~\cite{BoundedWidthJournal}: such algorithms  search for contradictions in an instance by examining the constraints on subsets of the variables of a fixed bounded size, and by  propagating this  information via bounded-size connections with other such subsets; they accept the instance if they cannot find any contradiction. For instance, $\CSP(\Q;<^\Q)$ from  Example~\ref{ex:acyclic} can be solved this way: the algorithm computes the transitive closure of the constraints by looking at 3-element subsets of the variables, and accepts if it does not produce a loop. The CSP in Example~\ref{ex:linear}, on the other hand,  cannot: Gaussian elimination is not local in this sense.

Since every computational problem can be encoded as the CSP of some template~\cite{BodirskyGrohe}, we cannot expect a uniform algebraic approach for all templates. However, the fundamentals of the algebraic approach for finite templates from~\cite{JBK}  lift almost verbatim to countably infinite ones (such as $(\mathbb Q;<^\Q)$) under the assumption that for each instance $I$ of the CSP, there are only finitely many different solutions up to automorphisms of the template~\cite{Topo-Birk}. To illustrate this  property, consider $\CSP(\mathbb Q;<^{\Q})$, and the instance $I$ with variables $x_1,x_2,x_3$ and constraints $x_1<x_2$, $x_1< x_3$. Clearly, $I$ has a solution $s$ setting $s(x_1)=0, s(x_2)=1, s(x_3)=1$. But $I$ also has infinitely many other solutions, most of which are however \emph{equivalent} in that they belong to the same \emph{orbit} of the template: an automorphism of $(\mathbb Q;<^\Q)$ sends one solution to another one. Up to orbit-equivalence, there are only three different solutions of this particular instance $I$,  which we can describe as follows:  $s(x_1)<^\Q s(x_2)<^\Q s(x_3)$, $s(x_1)<^\Q s(x_3)<^\Q s(x_2)$, and $s(x_1)<^\Q s(x_2)=^\Q s(x_3)$. Therefore, what we are looking for in this CSP is actually not rational numbers, but an orbit for the triple $(x_1,x_2,x_3)$ in the template as specified by  the order relation $<^\Q$ and the equalities which hold on it, such that  all constraints become true (which is possible if and only if the constraints do not imply that $s(x_i)<^\Q s(x_i)$ for any variable $x_i$). Countably infinite structures with the property that for every $n\geq 1$, there are only finitely many distinct $n$-tuples up to orbit-equivalence are called \emph{$\omega$-categorical}. Note that the ring of integers in Example~\ref{ex:diophantine} above is very much  not $\omega$-categorical: any two elements of $\mathbb Z$ are non-equivalent since the template has no automorphisms at all.

While  $\omega$-categoricity guarantees the availability of certain algebraic methods to investigate the mathematical structure of the template, it turns out that contrary to the finite-domain case the structure as measured by these methods is, without further assumptions,  largely insufficient to make  
predictions about the computational complexity of the CSP~\cite{BodirskyGrohe,GJKMP-conf, GJKMP}. In particular, even if the ``solution space" (i.e., all orbits of $n$-tuples) for every given instance $I$ (of length $n$) is finite, it might not be possible to algorithmically  enumerate all orbits of arbitrary length. In order to achieve this, it makes sense to first fix a simple way of describing orbits. Under $\omega$-categoricity, every  orbit can be defined by a single first-order formula (see~\cite{Hodges}); observe, however, that in the example of $(\mathbb Q;<^\Q)$, the orbit of any tuple is already completely determined by the relations that hold on it, and hence by a conjunction of atomic and negated atomic formulas. For a simple example where this is not the case, note that in the structure $(\mathbb Q;<^\Q,\{0\})$  the pairs $(-2,-1)$ and $(1,2)$ are in different orbits, but satisfy the same relations. Structures where orbits of tuples are determined by the  relations on them alone are called \emph{homogeneous}.

There is one more  ingredient for the class of  templates we will consider:  
we wish to be able to decide which orbits actually occur in the structure (otherwise the CSP can still be undecidable). For example, imposing the relations $s(x_1)<^\Q s(x_2)<^\Q s(x_3)<^\Q s(x_1)$ to describe the orbit of a potential solution $s$ to an instance of $\CSP(\Q;<^\Q)$ is invalid since no orbit of triples satisfies them. To provide a simple and in particular  polynomial-time (in fact: locally)  verifiable description of the orbits available as  solutions to an instance, the notion of \emph{finite boundedness}   seems natural: a homogeneous structure is finitely bounded if it has finitely many relations and there exists a finite set of ``forbidden conditions" given by conjunctions of atomic and negated atomic formulas such that the orbits which appear in the structure are described by precisely those tuples which do not realize any of the forbidden conditions. For example, in the case of $(\mathbb Q;<^\Q)$, such forbidden conditions would be   $\neg (a= b)\wedge  \neg(a<b)\wedge  \neg(b<a)$ (guaranteeing that any two distinct elements of a tuple are comparable in $<^\Q$); the condition $a<a$ (guaranteeing that no element of a tuple is related to itself in $<^\Q$); and the condition  $a<b\wedge b<c\wedge \neg(a<c)$ (guaranteeing that  $<^\Q$ is transitive on any tuple).

If a template $\A$ is  homogeneous and finitely bounded, then $\CSP(\A)$ is in NP: for any given instance $I$, there are only finitely many   non-equivalent potential solutions ($\omega$-categoricity, which readily  follows from  homogeneity and the  finite language); the elements of the  solution space (i.e., orbits) can be described by the relations which hold on their  tuples  (homogeneity), making it possible to guess one of the  orbits by guessing such relations;  whether or not  this guess actually yields an element of the solution space (i.e., a valid orbit) can be checked by local verification on subtuples of bounded size (finite boundedness); and finally, in the same fashion, so can whether the guess satisfies the constraints. We also see from this argument that each instance $I$ of the CSP of a finitely bounded homogeneous structure $\A$ whose maximal arity of relations is $k$ can be transformed into an equivalent instance $I'$ of a finite-domain CSP as follows: given variables $V=\{x_1,\ldots,x_n\}$ and a list $C$ of constraints, $I'$ has as variables all $k$-tuples of elements of $\{x_1,\ldots,x_n\}$, and the task is to assign to each such $k$-tuple one of the finitely many orbits in $\A$   in such a way that the assignment is  consistent (whenever $k$-tuples of variables intersect, the choice for their orbits must  agree on the intersection), that no forbidden condition is realized, and that each $k$-tuple satisfies the constraints in $C$. In the example of $(\Q;<^\Q)$, we have to pick for every pair $(x_i,x_j)$ one of the possibilities $s(x_i)<^\Q s(x_j)$, $s(x_j)<^\Q s(x_i)$, or $s(x_i)=^\Q s(x_j)$, in such a way that all constraints are met, that equality is transitive, and that we do not produce any of the forbidden subpatterns above. We remark that this translation from $I$ to $I'$ does not mean that the  problem $\CSP(\A)$ is equivalent to the CSP of a finite structure, since all finite-domain instances we produce as above  have non-trivial overlaps between their variables; in particular, a no-constraint instance is never obtained since already the transitivity of the equality relation imposes non-trivial constraints. Put differently, we have here a reduction to a finite-domain  CSP that might be harder than the original.

The finite-domain CSP instance $I'$ associated with an instance $I$ of $\CSP(\A)$ as above 
 is of a very particular form: basically, the constraints of $I'$ specify for each variable a set of allowed values among the orbits of $k$-tuples of $\A$, plus there are compatibility constraints which stem from the overlap of the variables and the forbidden conditions; but no other relations on the set of possible values (still  the set of $k$-orbits) are used. If we wish to allow such relations, then this amounts to considering $\CSP(\B)$ for structures $\B=(A;R_1^\B,\ldots,R_q^\B)$ whose relations are unions  of orbits of $\A$; we call such structures \emph{first-order reducts} of $\A$.  An example for  $\A=(\mathbb Q;<^\Q)$ would be $\B=(\mathbb Q;B^\Q)$, where $B^\Q$ is the ternary relation containing all $(a,b,c)$ such that $a<^\Q b<^\Q c$ or $c<^\Q b<^\Q a$ (hence $B^\Q$ is the union of two orbits); this yields the classical  NP-complete \emph{betweenness problem}~\cite{Opatrny}. In any instance $I$ of $\CSP(\sB)$, a constraint $B(x_i,x_j,x_k)$ corresponds to a $\binom{3}{2}$-ary constraint in the corresponding instance $I'$ whose variables are pairs of variables of $I$: that constraint states for the triple  $((x_i,x_j),(x_j,x_k),(x_i,x_k))$ of pairs that under a solution $s$ none of the pairs is sent to the equality orbit, that $s(x_i)<^\Q s(x_j)$ if and only if  $s(x_j)<^\Q s(x_k)$, and so forth.

For the same reasons as above $\CSP(\B)$ is in NP for every first-order reduct $\sB$ of a finitely bounded homogeneous structure $\A$. The following conjecture has been confirmed for the first-order reducts of various finitely bounded homogeneous structures including $(\Q;<^\Q)$\cite{tcsps-journal}, all homogeneous graphs~\cite{equiv-csps,BodPin-Schaefer-both, BMPP16}, the random partial order~\cite{posetCSP16}, the random tournament~\cite{MottetPinskerSmooth}, any unary structure~\cite{BodMot-Unary}, and various others~\cite{Phylo-Complexity,MMSNP-Journal,BodirskyBodorUIP,BodorDiss}. 
\begin{conjecture}[Bodirsky and Pinsker 2011; see~\cite{BPP-projective-homomorphisms}]\label{conj:BP}
Let $\B$ be a first-order reduct of a finitely bounded homogeneous structure. Then $\CSP(\B)$ is in P or NP-complete.
\end{conjecture}

\section{The Dilemmas}

\subsection{The first dilemma of the infinite sheep}\label{sect:first}

The algebraic approach to CSPs with finite or $\omega$-categorical template $\sA$ is based on the observation that   homomorphisms from finite powers of $\A$ into $\A$, called \emph{polymorphisms} of $\A$, preserve solutions: if $f(x_1,\ldots,x_\ell)$ is a polymorphism and $s_1,\ldots,s_\ell$ are solutions to an instance $I$ of $\CSP(\sA)$, then $f(s_1,\ldots,s_\ell)$ is a solution to $I$ as well. The set $\Pol(\sA)$ of all polymorphisms of $\sA$ can thus be viewed as symmetries of any solution set, and in fact $\Pol(\sA)$ determines the collection of all solution sets: any set of tuples which is invariant under all polymorphisms is a projection of the solution set  of some instance~\cite{JBK, BodirskyNesetrilJLC}. It then follows easily that the computational complexity of $\CSP(\sA)$ is determined by $\Pol(\sA)$, and  that the less polymorphisms $\sA$ has, the harder its CSP is since,  heuristically speaking, more unstructured search becomes necessary to solve it. 

As it turns out, in the case of a finite template $\sA$ even the symmetries of $\Pol(\sA)$ as measured by the non-nested \emph{identities} (i.e., universally quantified equations) that are witnessed in it determine the complexity of $\CSP(\sA)$~\cite{wonderland}. In particular, the theorems of Bulatov and Zhuk  state that $\CSP(\sA)$ is in P if and only if $\Pol(\sA)$ satisfies any non-trivial set of  non-nested identities  (\emph{non-trivial} meaning the identities are not satisfied by the polymorphisms of  all structures); this is the case if and only if $\sA$ has a $6$-ary polymorphism $s$ satisfying the \emph{Siggers identity} (see~\cite{Siggers})   $s(x,y,x,z,y,z)=s(y,x,z,x,z,y)$ (for all $x,y,z\in A$), and there are various  equivalent  conditions including the satisfaction of a \emph{cyclic identity}~\cite{Cyclic} or \emph{weak near-unanimity (wnu)  identities} for some arity~\cite{MarotiMcKenzie}. A  similar characterization exists for  bounded width   via wnu identities of \emph{all} arities $\geq 3$~\cite{BoundedWidthJournal} and some other identities~\cite{Maltsev-Cond}. For templates within the range of Conjecture~\ref{conj:BP}, it is believed that membership in P can be described by the local (i.e., on all finite subsets of the domain) satisfaction of non-trivial non-nested identities~\cite{wonderland,BKOPP,BKOPP-equations,GJKMP-conf,GJKMP}, since this condition obstructs the most obvious reason for NP-hardness, namely the \emph{pp-construction} of an NP-hard finite template. The condition  implies the global satisfaction of the  (slightly nested) \emph{pseudo-Siggers identity} $e\circ s(x,y,x,z,y,z)=f\circ s(y,x,z,x,z,y)$~\cite{BartoPinskerDichotomy, Topo}. 
This identity is a weaker condition than the various conditions known to characterize  polynomial-time tractability for finite domains. It neither implies a single non-trivial non-nested  identity (as witnessed by $(\Q;\neq^\Q,Z^\Q)$, where $Z^\Q$ is 4-ary and defined by the formula  $x_1\neq x_2\vee x_3=x_4$; here all  polymorphisms are injective up to dummy variables~\cite{BodChenPinsker}), nor any    \emph{pseudo-cyclic identity} $e\circ c(x_1,\ldots,x_\ell)=f\circ c(x_2,\ldots,x_\ell,x_1)$ of arity $\geq 2$ (as witnessed by  $(\Q;<^\Q,Z^\Q)$); 
it is also not known to imply \emph{pseudo-wnu identities} for some arity. This is at least in part due to the fact that $\omega$-categoricity of $\sA$ is incompatible with  \emph{idempotency} of $\Pol(\sA)$, a central  property for deriving  identities in finite (and even infinite~\cite{Pseudo-loop,olsak-idempotent}) algebras: idempotency means that the only unary polymorphism of $\sA$ is the identity function, and finite-domain CSPs can always be reduced to idempotent ones; $\omega$-categoricity, on the other hand, implies that $\sA$ has many unary polymorphisms, in fact many  automorphisms (since it has few orbits).
\begin{quote}
    \emph{``Idempotency: can't live with it, can't live without it."}
\end{quote}
An approximation of   idempotency in the $\omega$-categorical case is the notion  of a \emph{model-complete core}~\cite{Cores-journal}, which means that all unary polymorphisms are equal to some automorphism on every finite subset of the domain,  so in that sense there are as few unary polymorphisms as possible. If $\sB$ is a first-order reduct of a finitely bounded homogeneous structure which is \emph{Ramsey} (a combinatorial property, see~\cite{Topo-Dynamics}), then there is a model-complete core $\sB'$ with the same CSP as $\sB$ which is also a first-order reduct of a finitely bounded homogeneous Ramsey structure~\cite{MottetPinskerCores}. Hence, 
Conjecture~\ref{conj:BP} is equivalent to its restriction to model-complete cores in the case of first-order reducts of  finitely bounded homogeneous Ramsey structures. It is an open problem whether every first-order reduct of a finitely bounded homogeneous structure is also a first-order reduct of a finitely bounded homogeneous Ramsey structure. See~\cite{BPT-decidability-of-definability,Van-The-Survey, EvansHubickaNesetril} for more on this question and its variants.

\subsection{The third    dilemma of the infinite sheep}

We now fix a finitely bounded homogeneous structure $\sA$ 
of maximal arity $k$ and a first-order reduct $\sB$. 
If instances of  $\CSP(\sB)$ can be transformed into finite-domain CSP instances whose variables  take values in the orbits of $k$-tuples of $\sA$, and the  polymorphisms of $\sB$ determine the complexity of CSP($\sB$), then why does membership in P  have a description in terms of polymorphisms for finite templates, while Conjecture~\ref{conj:BP} is still open? The reason lies in the fact that the polymorphisms of $\sB$ need not act on the orbits of $\sA$, and hence do not translate into polymorphisms of the corresponding finite-domain CSP. In other words, the transformation of the instance $I$ of $\CSP(\sB)$ into a finite-domain CSP instance $I'$  might destroy the very symmetries of the solution sets we would like to use in order to prove membership in P.
\begin{quote}
  {\it ``Orbit equivalence: can't live with it, can't live without it."}
\end{quote}
 Those polymorphisms which do preserve orbit equivalence, and hence act naturally on the orbits of $\sA$, are called \emph{canonical with respect to $\sA$}~\cite{BPT-decidability-of-definability,RandomMinOps,BP-canonical}, and most of the research on Conjecture~\ref{conj:BP} has been based on this notion. If the canonical polymorphisms of $\sB$ satisfy non-trivial non-nested  identities in their action on orbits, then $\CSP(\sB)$ is in P thanks to the reduction  above~\cite{Bodirsky-Mottet}. Moreover,  bounded width is  implied by the  identities known from finite-domain CSPs when satisfied by canonical polymorphisms in their action on orbits, and the amount of locality needed in that case to solve the CSP is bounded by certain parameters which only depend on  $\sA$~\cite{Collapses}. 

Since we only know that  the entire polymorphism clone  $\Pol(\sB)$, and not just its canonical part,  determines the complexity of $\CSP(\sB)$,  the natural question then becomes how well the smaller set of  canonical polymorphisms represents $\Pol(\sB)$. It has been  observed in~\cite{BPT-decidability-of-definability} (see also~\cite{BP-reductsRamsey}), and exploited in all  classifications since~\cite{BodPin-Schaefer-both}, that if $\sA$ is not only finitely bounded and
homogeneous, but moreover  Ramsey, then there are many  canonical polymorphisms of $\sB$ with respect to $\sA$ in the sense that every polymorphism of $\sB$ \emph{locally interpolates} a canonical one modulo the orbits of $\sA$: for every $f(x_1,\ldots,x_\ell)\in\Pol(\sB)$, and for all orbits $O_1,\ldots,O_\ell$  of $\sA$, there exist tuples $t_1,\ldots,t_\ell$ in these orbits such that $f$ is on $t_1,\ldots,t_\ell$  the restriction of a canonical function. 

While the ubiquity of canonical polymorphisms in the sense above seems  encouraging, it does not guarantee the satisfaction of non-trivial identities by the canonical polymorphisms if the  other polymorphisms do. Nonetheless, one successful line of research has been to count on  exactly that, leading in particular to  proofs of  P/NP-complete dichotomies  for the  CSPs of the first-order reducts of any homogeneous graph~\cite{BMPP16} and various other classes of  CSPs~\cite{posetCSP16,BodMot-Unary, MMSNP-Journal,MottetPinskerSmooth,BodirskyBodorUIP,BodorDiss}; also for characterizing bounded width this approach has been fruitful~\cite{MottetPinskerSmooth,Collapses}. A systematic approach to  comparing the set of canonical polymorphisms with that of all polymorphisms was developed  in~\cite{MottetPinskerSmooth}, based on so-called \emph{smooth approximations}, in many cases deriving a contradiction from the assumption that the canonical polymorphisms do not witness any non-trivial identities while the other polymorphisms do. 

Observe  that in the cases where membership in P is witnessed by canonical polymorphisms, the only algorithm employed to show polynomial-time tractability is the one from finite-domain CSPs, 
used as a blackbox; in other words, the CSP is simply reduced to a finite one which is then solved by the   hands of Bulatov and Zhuk. A similar  phenomenon occurs whenever bounded width of $\CSP(\sB)$ is witnessed by canonical polymorphisms with respect to $\sA$: 
the CSP can then essentially be solved by locally checking a finite-domain CSP instance, resulting in a required amount of locality which is bounded by what is needed for the finite-domain instance and an overhead from the reduction which depends only on $\sA$.

\subsection{The second dilemma of the infinite sheep}

Powerful as the approach of a blackbox reduction to finite-domain CSPs via canonical functions might have turned out in many  cases, the sword of Damocles has been hanging over the heads of those who master its technicalities ever since the first grand complexity classification for CSPs of structures within the range of Conjecture~\ref{conj:BP}, namely the first-order reducts of $(\Q;<^\Q)$ in~\cite{tcsps-journal}. For such CSPs,   polynomial-time tractability  is actually never witnessed by canonical polymorphisms with respect to $(\Q;<^\Q)$ (except in trivial cases). A simple example is the non-canonical polymorphism $(x,y)\mapsto \max(x,y)$: from the information $x<^\Q x'$ and $y>^\Q y'$, we do not know whether $\max(x,y)<^\Q\max(x',y')$; hence the function $\max$ does not act on  orbits of pairs in $(\Q;<^\Q)$. On the other hand, having this polymorphism, which is in particular a polymorphism of the structure $(\Q;<^\Q)$ itself,  implies polynomial-time solvability of the CSP~\cite{tcsps-journal} (see also the reduction to a finite-domain CSP  in~\cite{MottetPinskerSmooth} which avoids canonical functions).

Unfortunately, the order $(\Q;<^\Q)$ is not an  exception, but at the heart of a dilemma inherent even in the study of CSPs  of first-order reducts $\sB$ of finitely bounded homogeneous structures $\sA$ which have, a priori, no connection with an order. 
 The reason is that the local interpolation of  polymorphisms which are canonical with respect to $\sA$ by arbitrary polymorphisms of $\sB$ as  mentioned above is closely connected to (and in some sense equivalent to, see~\cite{BP-canonical}) $\sA$ being Ramsey. 
 The Ramsey property is equivalent to the automorphism group of $\sA$ being \emph{extremely amenable}~\cite{Topo-Dynamics}, which in turn  implies that the automorphisms of $\sA$ leave some linear order $<$ on the domain of $\sA$ invariant. Hence, canonical functions with respect to $\sA$ are canonical with respect to the expansion $(\sA,<)$ of $\sA$ by the order $<$, since canonicity is defined purely in terms of the automorphisms of $\sA$.  
 It follows that those   polymorphisms of $\sB$ which are canonical with respect to $\sA$  cannot satisfy any non-trivial non-nested identities in their action on orbits, at least if $\sB$ is a model-complete core, for similar reasons as is the case with the first-order reducts of $(\Q;<^\Q)$. Whence, the canonical polymorphisms of $\sB$ with respect to $\sA$ cannot witness polynomial-time tractability of its CSP.
 
 We outline the proof of our claim that in the above situation, the canonical polymorphisms do not satisfy any non-trivial non-nested identities in their action on orbits. Otherwise, their action on orbits would contain, for some $l\geq 2$, a  function $f$ satisfying the 
 \emph{cyclic identity}  $f(x_1,\ldots,x_\ell)=f(x_2,\ldots,x_\ell,x_1)$ by~\cite{Cyclic}. By Ramsey's theorem, there exists an infinite $S\subseteq A$ which is \emph{order indiscernible}, i.e., two tuples on $S$ belong to the same orbit if and only if the order $<$ agrees on them. 
  The cyclic identity of $f$ on orbits then implies $f(a_1,\ldots,a_\ell)=f(a_2,\ldots,a_\ell,a_1)$ for all $a_1,\ldots,a_\ell\in S$, since otherwise $<$ would relate some element to itself. Using this fact for any  $a_1,\ldots,a_\ell\in S$ with $a_1<\cdots<a_\ell$ yields that on $S$,  $f(b_1,\ldots,b_\ell)=f(c_1,\ldots,c_\ell)$ whenever $b_1<c_1,\ldots,b_{\ell-1}<c_{\ell -1}$ and $c_\ell<b_\ell$. From this it follows that $f$ is constant on $S$, contradicting the assumption that $\sB$ is a model-complete core. 
\begin{quote}
\emph{``Extreme amenability: can't live with it, can't live without it."}\hfill (A. Mottet)
\end{quote}
Given this argument, how can canonical functions be of use at all, and how were they used in the mentioned  classifications? Let us consider, for example, the random graph $\G$ and a  first-order reduct $\B$: both  polynomial-time tractability of $\CSP(\sB)$ and its solvability by local consistency checking have been characterized by canonical polymorphisms with respect to $\G$~\cite{BodPin-Schaefer-both,MottetPinskerSmooth}. Yet,  $\G$ is not a Ramsey structure; in particular, the polymorphisms of $\B$ do not necessarily locally  interpolate canonical functions with respect to $\G$. Both the original approach in~\cite{BodPin-Schaefer-both} as well as the more recent and systematic  approach in~\cite{MottetPinskerSmooth} expand $\G$ by a linear order $<$ on its domain in such a way that the expanded structure $(\G,<)$ is a finitely bounded homogeneous Ramsey structure, and consider $\B$ as a first-order reduct of $(\G,<)$. It is an open problem whether such expansions  exist for all finitely bounded homogeneous structures $\sA$ (in general, adding a linear order is not sufficient, but a finite number of relations including a linear order could be); see  the remark at the end of Section~\ref{sect:first}. It then follows  that every  polymorphism of $\B$ locally interpolates a canonical polymorphism with respect to $(\G,<)$; the set of such canonical polymorphisms does not, however, satisfy any non-trivial non-nested~identities in its action on orbits of $(\G,<)$, by the argument above. To escape this dilemma, in both approaches canonical polymorphisms with respect to $(\G,<)$ are composed 
to obtain  polymorphisms which are canonical with respect to $\G$. Finally,  both approaches find such polymorphisms which additionally  satisfy non-trivial non-nested identities in their action on orbits of $\G$ whenever $\B$ has a pseudo-Siggers polymorphism (the approach in~\cite{BodPin-Schaefer-both} does not yet use this algebraic fact but makes an exhaustive case distinction over all possible canonical functions with respect to $(\G,<)$). Hence, polynomial-time  tractability of $\CSP(\sB)$ follows; the approach in~\cite{MottetPinskerSmooth}
 for bounded width proceeds similarly.

Naturally, one then wonders under what conditions it is possible to compose polymorphisms which are canonical with respect to a Ramsey expansion $\sA'$ of a finitely bounded homogeneous structure $\sA$ to obtain  polymorphisms which are canonical with respect to $\sA$, or in other words, whose action on orbits does not depend on the order $<$ anymore. It was remarked  in~\cite{BodirskyBodorUIP} that the dividing line between those structures $\sB$ where polynomial-time tractability of the CSP is witnessed by canonical functions, and those where this is not the case, empirically corresponds to the \emph{strict order property (SOP)}  (structures with the SOP such as $(\Q;<^\Q)$ falling into the latter class). The SOP (see~\cite{Simon}) states about a structure   that there is a  formula in its  theory  which defines a preorder with infinite chains. 
The mentioned remark is somewhat vague in that it does not specify  whether the SOP is considered for $\sB$ or $\sA$ (the  Ramsey expansion $\sA'$ always has the SOP), and whether canonicity is meant with respect to $\sA$ or $\sA'$, but in any case does not seem accurate: in particular, the random poset $\sP$ has the SOP, and membership in P is  described by canonical polymorphisms with respect to $\sP$ for its first-order reducts which are model-complete cores. Along similar lines, in~\cite[Conjecture 6.2]{BodorDiss} it is conjectured that if $\sB$ does not have the SOP, then the existence of a pseudo-Siggers polymorphism of $\sB$ implies the existence of such a polymorphism which is  canonical with respect to $\sA'$ (and where $e,f$ in  the pseudo-Siggers identity are from $\Pol(\sA')$). This seems to be contradicted already by the  first-order reduct $(\Q;\neq^\Q,Z^\Q)$ of 
$(\Q;<^\Q)$ with $Z^\Q$ as in Section~\ref{sect:first}. 

A statement which seemed conceivable to the author  until recently was  that whenever $\sA$ does not have the SOP, then polynomial-time tractability of $\CSP(\sB)$ is witnessed by canonical polymorphisms with respect to $\sA$ (but the converse does not hold). We will give a counterexample below.

\section{Opening the Blackboxes}\label{sect:blackboxes}

We now provide two examples exhibiting some limitations of the blackbox use via canonical functions of the  finite-domain  algorithms to  solve a CSP in polynomial time or check its local consistency. The first example, a result of discussions with Mottet, is a first-order reduct of the random 3-hypergraph $\sH$ (which does not have the SOP) which is a model-complete core and has a pseudo-Siggers polymorphism, and hence its CSP should be in P according to the detailed version of Conjecture~\ref{conj:BP}; it does, however, not have any canonical polymorphisms with respect to $\sH$  witnessing this. This example can either be implemented so that its CSP has, in spite of this,  bounded width, or so that it does not. 

The second example, from discussions with Mottet and Nagy, is a unary structure $\sA$ with the following property: every first-order reduct $\sB$ of $\sA$ whose CSP has bounded width can be solved by the blackbox use of the finite-domain local consistency algorithm via canonical functions, but doing so results in the use of an unnecessarily large amount of locality; more precisely, applying a local consistency algorithm directly one needs  less locality.

\subsection{Example 1: Hypergraphs}

Let $(\sH,<)=(H;E,<)$ be the random 3-hypergraph with a random order expansion; that is, it is the up to isomorphism unique homogeneous  structure   with the following properties:  
$E$ is a ternary totally symmetric relation containing only injective tuples,  $<$ is a linear order, and there are no additional forbidden conditions for the orbits. $(\sH,<)$ then can be viewed as an expansion of the structure $\sH=(H;E)$, the random 3-hypergraph, by a random linear order $<$. Both $(\sH,<)$ and $\sH$ are finitely bounded and homogeneous, and the first is Ramsey while the latter is not; in fact, $\sH$ does not have the SOP.

We first define a binary injection $f$ on $H$ which is canonical with respect to $(\sH,<)$ by specifying its action on orbits of triples; $f$ will not be canonical with respect to $\sH$.  Independently of the structure, we say that an orbit is injective if the tuples in it are; similarly, we call an orbit constant if the tuples in it are. There are precisely two injective orbits with respect to $\sH$: one is given by the relation $E$, and the other one,  consisting of the injective triples not in $E$, we call $N$. Each of the orbits $E$ and $N$ splits into $3!$ orbits of $(\sH,<)$ by the specification of a linear order. We first define the action of $f$ on orbits of pairs, of which there are three, given by the possibilities $x<y$, $x>y$, $x=y$. If $O_1,O_2$ are such orbits, then: if both $O_1,O_2$ are injective, then $f$ returns $O_1$; if both are constant, then so is their value under $f$ (since $f$ is a function); if precisely one of them is constant, then $f$ returns the other one. We next specify the action of $f$ on orbits $O_1,O_2$ of triples; here we need to be consistent with the specification on pairs.  Note that the order and the equalities on the orbit $f(O_1,O_2)$ are  already determined by the above, so we only have to specify whether $f(O_1,O_2)$ belongs to $E$ or to $N$ in case it is injective; this is the case if and only if $O_1$ and $O_2$ have no equality holding at the same position.  
If $O_1,O_2$ are both injective, then  $f(O_1,O_2)$ belongs to the same $\sH$-orbit as $O_1$. If precisely one of $O_1,O_2$ is injective, then $f(O_1,O_2)$ belongs to the same $\sH$-orbit as that injective orbit. 
The only remaining case is when $O_1,O_2$ are neither injective nor constant, and have their unique equality at different positions; we then define  $f(O_1,O_2)$ to be a suborbit of $E$ if and only if $O_1, O_2$ agree on whether the element which appears twice in their triples is smaller or larger than the third element with respect to $<$. Note that this makes $f$ non-canonical with respect to $\sH$. A function $f$ on $H$ with this action on orbits exists since our definition does not force any forbidden subpatterns to be realized in  $(\sH,<)$.

Let $m$ be a ternary function on $H$ which acts like a \emph{majority operation} on the orbits $E,N$ of $\sH$, i.e., it satisfies  $m(E,E,N)=m(E,N,E)=m(N,E,E)=m(E,E,E)=E$, and the dual equations where $E$ and $N$ are flipped. We moreover require  $m(O_1,O_2,O_3)=O_1$ for all injective orbits $O_1,O_2,O_3$ of pairs in $(\sH,<)$. 
This is not a complete specification of an  action on orbits of $(\sH,<)$, but we will not require more information in what follows.
Set $h(x,y,z)=f(x,f(y,z))$ and 
$$
g(x,y,z)=m(h(x,y,z),h(y,z,x),h(z,x,y))\; .
$$
Then $g$ is canonical with respect to $(\sH,<)$ but not $\sH$. If $s$ is any binary injective function obtained by composing $g$ and automorphisms of $\sH$ which are canonical with respect to $(\sH,<)$, then $s$ is still not canonical with respect to $\sH$: the canonical  automorphisms are either increasing or decreasing with respect to $<$, and an easy induction shows that flipping the order in one  argument of $s$ when acting on $(\sH,<)$ can change its value from a suborbit of $E$ to a suborbit of $N$ (this is inherited from $f$). If we allow arbitrary  automorphisms of $\sH$ in the composition, then $s$ is still not canonical with respect to $\sH$: roughly, the argument is that  such automorphisms locally interpolate  automorphisms which are canonical with respect to  $(\sH,<)$.  

The set of all relations on $H$ which are invariant under all automorphisms of $\sH$ as well as under $g$ then gives rise to a first-order reduct $\sB$ of $\sH$, since all such  relations are, by definition, unions of orbits of $\sH$. The polymorphisms of $\sB$ contain $g$ by definition, and all polymorphisms $f(x_1,\ldots,x_\ell)$ with $\ell\geq 2$ which depend on all of their variables are not canonical with respect to $\sH$: by the general basic theory of polymorphisms on infinite sets (see e.g.~\cite{GoldsternPinsker}) any such $f$ is  injective, and  $f(x,y,\ldots,y)$ has the property of $s$ above. By the same argument, we see that $\sB$ is a model-complete core.  The structure $\sB$ has an infinite number of relations, but in order to obtain  a well-defined CSP we can replace $\sB$ by a suitable finite subset of its relations while maintaining the mentioned properties: there are only finitely many actions of canonical functions on orbits of triples of $\sH$, so all these possibilities can be excluded that way.

Clearly, for all tuples $t_1,t_2,t_3$ of elements in $H$, the tuples $g(t_1,t_2,t_3),g(t_2,t_3,t_1)$  belong to the same orbit with respect to $\sH$. A standard compactness argument then yields that $g$ witnesses the ternary    pseudo-cyclic  identity~\cite{BPP-projective-homomorphisms}; consequently, $\sB$ also has a  pseudo-Siggers polymorphism~\cite{BartoPinskerDichotomy, Topo}.

The problem  $\CSP(\sB)$ is in P  since it has bounded width:   yet unpublished  arguments by Mottet, Nagy, and the author use the existence of an \emph{absorbing subuniverse} on the set of 
triples, namely the injective triples as witnessed by injectivity of $g$, and solvability of instances where all triples are  constraint by this subuniverse by  local consistency checking; the latter is implied   by the action of $g$ as a majority operation on the orbits $E,N$  of this  subuniverse, and reduction to the corresponding finite-domain instance.

The function $g$ can, however, easily be modified to obstruct the applicability of local consistency algorithms by choosing $m$ to act like a minority rather than a majority operation on $\{E,N\}$. In that case, $\CSP(\sB)$ should still be in P since $\sB$ still has a pseudo-Siggers polymorphism.  One approach for proving membership in P could be  to lift Zhuk's reduction to absorbing subuniverses for finite-domain CSPs into this context.

\subsection{Example 2: Unary structures}
A \emph{unary structure} is a relational structure which has only unary relations. The study of the CSPs of first-order reducts of such structures reduces to that of CSPs of structures of the form $\B=(A;A_1^\sA,\ldots,A_m^\A,R_1^\B,\ldots,R_q^\B)$, where $A_1^\A,\ldots,A_m^\A$ are subsets which form a partition of $A$, and $R_1^\B,\ldots,R_q^\B$ are unions of orbits of the structure $\sA=(A;A_1^\A,\ldots,A_m^\A)$; in other words, we are looking at \emph{first-order expansions} of $\sA$. Let us focus here on the case where all parts $A_r^\A$ of the partition are infinite.   In a solution $s$ to an instance $I$ of $\CSP(\sB)$ with variables $\{x_1,\ldots,x_n\}$, in order to choose an orbit for $(s(x_1),\ldots,s(x_n))$ in $\sA$ we have to pick for every variable $x_i$ one part $A_r^\A$ of the partition into which $s$ maps that variable, and whenever two variables are mapped into the same part, we have to choose whether to give these two variables equal value or not; these two decisions  together completely determine the orbit of $s$ in $\sA$. By the infinity of the parts, there is enough space to give different values to variables $x_i,x_j$ whenever we wish to even if they are mapped to the same part, so any  assignment as above actually defines an orbit provided the distributed equalities   satisfy transitivity.  The constraints of the instance $I$ are statements of the form $A_r(x_i)$ and $x_i=x_j$ and certain Boolean combinations of such statements as defined by constraints from $R_1,\ldots,R_q$.  Hence, this is essentially a finite-domain CSP (with possible values $a_1,\ldots,a_m$ corresponding to the parts $A_1^\A,\ldots,A_m^\A$ of the partition) with a twist involving the equality relation. We remark that this twist is non-trivial since  even constraints from the equality relation alone can result in an NP-hard CSP: for example, if $m=1$ and the partition has only one class, the CSP of the structure $(A;A_1^\sA,\neq^\sA,R^\sB)$, where $R^\sB$ is the 4-ary relation defined by the formula $(x_1= x_2)\vee  (x_3=x_4)$ is NP-complete. CSPs where the partition is trivial in this sense and which therefore only concern the equality relation on an infinite set are called \emph{equality CSPs}, and their complexity classification is simple: either the template has a polymorphism which is a  binary injection or a constant function, and the CSP has bounded width, or all polymorphisms depend injectively on precisely one variable, and the CSP is NP-complete~\cite{ecsps}.

Both membership in P~\cite{BodMot-Unary} and bounded width~\cite{Collapses} are characterized by canonical functions for the CSP of  structures $\B=(A;A_1^\sA,\ldots,A_m^\A,R_1^\B,\ldots,R_q^\B)$ as above. We now consider the latter situation: this is the case if and only if there are polymorphisms of all arities $\geq 3$ which are canonical with respect to $\sA$ and which  satisfy the wnu  identities in their action on orbits of $\sA$.

How much locality is needed to solve a given CSP instance $I$ in this case? If  we translate $I$ into a finite-domain instance $I'$, then $I'$ can be solved correctly by the \emph{$(2,3)$-minimality algorithm} which, roughly speaking, propagates the constraints of $I'$ on the pairs of its variables via triples. As a guiding example for the meaning of the parameters $(2,3)$, note in order to compute the transitive closure of a binary relation, we look on which pairs the relation holds (hence the parameter 2), and then consider triples (hence the parameter 3) in order to conclude on which other pairs it needs to hold in order to be transitive; we then repeat. That the (2,3)-minimality algorithm  is sufficient to correctly solve any instance of a finite-domain CSP that can be solved by local consistency checking, or more precisely the $(a,b)$-minimality algorithm for some $a,b\geq 1$, 
is the statement of the collapse of the bounded width hierarchy for finite-domain CSPs~\cite{BartoWidth}. A general argument examining the  reduction sending $I$ to $I'$ then shows that since the variables of $I'$ are pairs of variables of $I$, it is sufficient to  run $(4,6)$-minimality in order to solve $\CSP(\sB)$ ($2\cdot (2,3)=(4,6)$)~\cite{Collapses}.

Is this optimal? Considering that CSPs of structures $\sB$ as above are really  finite-domain CSPs combined with equality CSPs, that $(2,3)$-minimality solves every finite-domain CSP with bounded width, and that likewise  any equality CSP which is not NP-complete can be solved by  $(2,3)$-minimality (basically, any such CSP is solved by computing the transitive closure of equality), it seems reasonable to conjecture  that $(2,3)$-minimality is  sufficient also in the case of $\CSP(\sB)$. Arguments by Mottet and Nagy confirm this for partitions with two parts ($m=2$).

\section{Conclusion}
CSPs of first-order reducts of finitely bounded homogeneous structures  constitute a vast extension of finite-domain CSPs and can model additional natural  computational  problems  such as acyclicity of graphs. They  share many properties with finite-domain CSPs; in particular, every instance $I$ to such a CSP can  naturally be  reduced to a finite-domain instance $I'$. 
Powerful and elegant methods to obtain complexity classifications via this reduction have been developed around the notion of canonical polymorphism,  which is a certain symmetry of the solution sets of the CSP which survives the translation of the instance $I$ to $I'$. On the other hand, our examples show that this  reduction will not be sufficient for proving a general P/NP-complete dichotomy as in Conjecture~\ref{conj:BP} even under the failure of the  SOP; it also does not, in general, faithfully reflect  the power of local consistency algorithms. Opening the blackboxes of the algorithms for finite-domain CSPs and applying suitable adaptations directly seems to be required now in order to advance our understanding of the complexity of such CSPs, and promises an escape from the current dilemmas for all infinite sheep.

\bibliography{global.bib}
\bibliographystyle{plain}
\end{document}